\documentclass[aps,prl,twocolumn,unsortedaddress,superscriptaddress]{revtex4-2}

\usepackage{graphicx, hyperref, amsmath, amssymb, comment}
\usepackage[utf8x]{inputenc}
\usepackage{physics}
\usepackage{mathrsfs}
\usepackage{todonotes}
\providecommand{\hlf}[1][1]{\ensuremath{\frac{#1}{2}}}
\providecommand{\ort}[1][1]{\frac{#1}{\sqrt{2}}}

\providecommand{\fr}{(\vb{r})}

\DeclareMathOperator{\naturals}{\mathbb{N}}

\providecommand{\ii}{\ensuremath{i}}
\providecommand{\e}[1]{e^{#1}}
\providecommand{\fourier}{\mathcal{F}}
\providecommand{\unitary}{\hat{U}}

\providecommand{\azimuth}{\phi}

\providecommand{\Ev}[1][]{\vb{E}_{\text{#1}}}
\providecommand{\Es}[1][]{\ensuremath{E_{\text{#1}}}}
\providecommand{\Eamp}{\ensuremath{\mathcal{E}}}
\providecommand{\Bv}[1][]{\vb{B}_{\text{#1}}}
\providecommand{\Bs}[1][]{\ensuremath{B_{0\text{#1}}}}

\providecommand{\inten}[1][]{\ensuremath{\mathcal{I}}_\text{#1}}

\providecommand{\oam}{\ell \azimuth}
\providecommand{\ioam}{\ii \ell \azimuth}
\providecommand{\LC}{\ensuremath{\vu*{ \sigma }_{+}}}
\providecommand{\RC}{\ensuremath{\vu*{ \sigma }_{-}}}

\providecommand{\polarB}{\ensuremath{\theta_{\rm B}}}
\providecommand{\azimuthB}{\ensuremath{\phi_{\rm B}}}
\providecommand{\hamiltonian}[1][]{\ensuremath{\hat{H}}_{\text{#1}}}
\providecommand{\wavF}[1][]{\psi_{\text{#1}}}
\providecommand{\hc}{\text{H.c.}}

\providecommand{\volume}{\mathcal{V}}
\providecommand{\numberDensity}{\ensuremath{N_{\volume}}}

\providecommand{\rabi}[1][]{\ensuremath{\Omega_{\text{R#1}}}}

\providecommand{\larmor}[1][]{\ensuremath{\Omega_{\text{L#1}}}}
\providecommand{\adm}[1][]{\ensuremath{\hat{\rho}_{\text{#1}}}}

\providecommand{\decay}{\Gamma}
\providecommand{\decayop}{\hat{\decay}}
\providecommand{\repop}{\Lambda}
\providecommand{\repopop}{\hat{\repop}}

\providecommand{\energy}{U}
\providecommand{\normFacAngle}{N(\phi)}
\providecommand{\fourieraz}{\fourier_{\azimuth}}
\providecommand{\tp}{T_{1\rightarrow e}} 

\date{\today}
\hypersetup{
 pdfauthor={},
 pdftitle={},
 pdfkeywords={},
 pdfsubject={},
 pdfcreator={Emacs 26.1 (Org mode 9.1.14)}, 
 pdflang={English}}
\begin{document}

\title{An atomic compass -- detecting 3D magnetic field alignment with vector vortex light}

\author{Francesco Castellucci}
\thanks{F.C. and T.W.C. contributed equally to this work.}
\affiliation{School of Physics and Astronomy, University of Glasgow, G12 8QQ, United Kingdom}
\author{Thomas W. Clark}
\email{thomas.clark@wigner.hu}
\affiliation{Wigner Research Centre for Physics, Hungarian Academy of Sciences, H-1525, Hungary}
\author{Adam Selyem}
\affiliation{Fraunhofer Centre for Applied Photonics, G1 1RD, United Kingdom}
\author{Jinwen Wang}
\affiliation{School of Physics and Astronomy, University of Glasgow, G12 8QQ, United Kingdom}
\affiliation{Shaanxi Province Key Laboratory of Quantum Information and Quantum Optoelectronic Devices, School of Physics, Xi'an Jiaotong University, Xi'an 710049, China}
\author{Sonja Franke-Arnold}
\email{sonja.franke-arnold@glasgow.ac.uk}
\affiliation{School of Physics and Astronomy, University of Glasgow, G12 8QQ, United Kingdom}

\begin{abstract}
We describe and  demonstrate how 3D magnetic field alignment can be inferred from single absorption images of an atomic cloud.  While optically pumped magnetometers conventionally rely on temporal measurement of the Larmor precession of atomic dipoles, here a cold atomic vapour provides a spatial interface between vector light and external magnetic fields. Using a vector vortex beam, we inscribe structured atomic spin polarisation in a cloud of cold rubidium atoms, and record images of the resulting absorption patterns. The polar angle of an external magnetic field can be deduced with spatial Fourier analysis. This effect presents an alternative concept for detecting magnetic vector fields, and demonstrates, more generally, how introducing spatial phases between atomic energy levels can translate transient effects to the spatial domain.
\end{abstract}

\maketitle

Most investigations and applications of light-atom interaction are concerned with homogeneously polarized light, or scalar light. Light-atom interaction however, by its very nature, is a vectorial process, that depends explicitly on the alignment between an external magnetic field and the optical and atomic polarizations \cite{budker2002resonant,franke2017optical, labeyrie2018collective,babiker2018atoms,wang2020vectorial,ackemann2021collective}. 
Over the last decades, the generation and use of vectorial light fields with spatially varying polarization profiles has matured into an active research area, with a plethora of applications in the optical domain  \cite{zhan2009cylindrical,rubinsztein2016roadmap,rosales2018review,chen2018vectorial,forbes2019quantum}, including communication \cite{ndagano2017creation}, polarimetry  \cite{hawley2019passive} and super-resolution imaging \cite{dorn2003sharper}.
Our ability to design complex vector light fields now allows the full exploration of vectorial light-matter interaction \cite{wang2020vectorial}. 
One of the earliest examples is the prediction \cite{allen1994} and measurement \cite{barreiro2006} of the rotational Doppler effect, with more recent applications including complex image memories \cite{parigi2015storage,ye2019experimental}, manipulation of non-linear effects \cite{bouchard2016polarization,hu2019nonlinear}, investigations of spatial anisotropy \cite{fatemi2011cylindrical,wang2018optically,wang2019directly},  and spatially dependent electromagnetically induced transparency (EIT) \cite{radwell2015spatially,hamedi2018,yang2019observing}.  

Here, we investigate the role of external magnetic fields on the propagation of vectorial light fields through atomic gasses, and specifically demonstrate that the 3D alignment of a magnetic field can be deduced from a single absorption profile of a vector vortex beam.
Atomic gasses are optically active media with a highly sensitive external field response, making them ideal candidates for magnetometry
\cite{budker2000sensitive,kominis2003subfemtotesla,budker2007optical,grewal2020light}. 
Atomic magnetometers have been developed to detect magnetic gradients \cite{affolderbach2002all}, multiple components of the magnetic vector field \cite{yudin2010vector,patton2014all,thiele2018self,Ingleby2018,pyragius2019voigt}, or to compensate magnetic backgrounds in 3D \cite{smith2011three}. Typically, optically pumped atomic magnetometers 
are based on observing the coherent Larmor precession of polarized atomic spins in a magnetic field, whereas vector magnetometers may employ radio-frequency modulation to map the vector components onto different harmonics. 

\begin{figure}[tb]
  \centering
  \includegraphics[width=\linewidth]{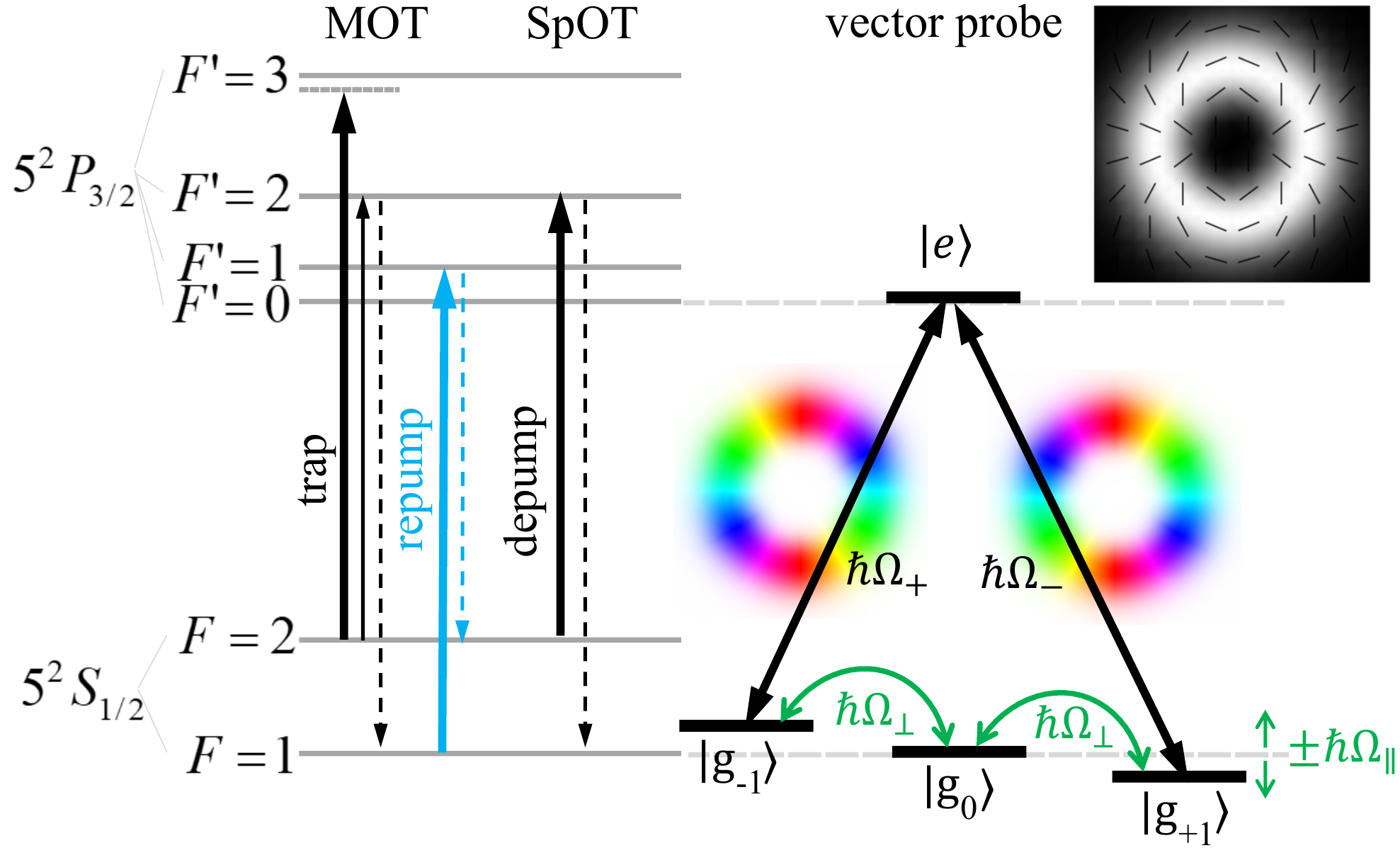}
  \caption{\label{fig:overview}
   Schematic energy levels and laser transitions: Rb$^{87}$ atoms are cooled and trapped in a standard MOT and then transferred into a SpOT, populating the $F=1$ ground state.  A vector vortex beam, drives a $\Lambda$-transition, where the $\sigma_{\pm}$ transitions carry opposite phase profiles and an external magnetic field couples the ground states. The phase profiles of the probe light are shown for $\ell=\pm 2$, with hue representing values between $0$ and $2\pi$. The top inset shows the corresponding intensity and polarization profile.
   }
\end{figure}

In this letter, we demonstrate a fundamentally different approach, replacing the \emph{dynamic} detection 
of the spin precession with the \emph{spatially resolved} detection of the atomic response to vector vortex light.  
We investigate the interaction of cold $^{87}{\rm Rb}$ atoms with vector vortex beams on the $D_2$ ($780\;$nm) $F=1 \to F'=0$ transition, as indicated in Fig.~\ref{fig:overview}, and show that the spatial transmission profile of such light depends strongly on the 3D alignment of a static external magnetic field. By observing the atoms' absorption profile, and specifically its Fourier decomposition, we can deduce the alignment of the magnetic field in three dimensions. 
Similar to other recent work \cite{chen2020calibration,Qiu2020Visualization}, our scheme requires only a single probe beam, thereby avoiding potential transverse dephasing effects. Unlike these previous schemes, the simple $F=1 \to F'=0$ configuration allows us to decouple the effect of 3D alignment from a modification of the magnetic field strength: demonstrating an atomic compass based on the absorption profile of a vector vortex beam.

Although our present demonstration uses a simple vector vortex beam, the principle applies to arbitrary vector light fields, with possible applications to inertial, gradient and position sensing, long term magnetic effects, as well as magnetic anomaly detection. 


\paragraph{Concept and theoretical model:}
The interaction of atoms with light is, to first order, determined by the atomic dipole Hamiltonian,
\begin{equation}  
\label{eq:H}
\hat{H}=-\mathbf{D}\cdot\mathbf{E} +g_F \mu_B\mathbf{F} \cdot \mathbf{B},
\end{equation} 
where $\mathbf{E}$ and $\mathbf{B}$ are the electric vector field and the external static magnetic field; $\mathbf{D}$ and $\mathbf{F}$ the induced atomic electric dipole and atomic spin polarization; $g_F$ the Landé g-factor and $\mu_B$ the Bohr magneton. 
In a closed system, an equilibrium can be reached, where the steady-state atomic system is polarized according to the optical polarization pattern, and the response of the optical field depends on the alignment of the optical polarization with respect to $\mathbf{B}$. 

Circularly polarized light generates atomic dipole moments, causing optical dichroism, whereas linear polarization leads to atomic quadrupole moments, which generate birefringence \cite{auzinsh2010optically}. A vector vortex beam,
\begin{equation}   
\label{eq:Efield}
     \Ev\fr = \ort \Es (r)\left[ \e{-\ioam} \LC + \e{\ioam}\RC \right],
\end{equation}
represents the latter case. Here, the left and right circularly polarized components, $\vu*{ \sigma }_{\pm}$, carry equal and opposite orbital angular momentum (OAM), $\mp \ell$, resulting in a polarization pattern, whose linear polarization rotates with the azimuth, $\phi$, as shown in the inset of Fig.~\ref{fig:overview}.
Although such absolute phase effects are generally meaningless, a break in symmetry, \textit{e.g.}~due to an external magnetic field, can make the dependence measurable \cite{radwell2015spatially}. 

The inscribed structure of magnetic quadrupole moments generates locally varying birefringence, in turn modifying the propagation of the light through the atomic sample. The induced atomic alignment precesses around an applied magnetic field,
\begin{equation}         
\label{eq:Bfield}
     \Bv\fr = \Bs \left( \sin\polarB \cos\azimuthB \vu{x} + \sin\polarB \sin\azimuthB \vu{y} + \cos\polarB \vu{z} \right),
\end{equation}
where $\polarB$ and $\azimuthB$ denote the inclination from the propagation axis and the azimuthal angle, respectively.

The atomic response is determined by the interplay between the local polarization direction of the light and the global external magnetic field.
The spatial features of the resulting absorption profile can be analysed in terms of their angular Fourier decomposition, allowing us to identify the 3D magnetic field alignment from a single absorption image.

We consider a standard Zeeman, \(\Lambda\)-type transition, resonantly coupling the $F=1,\; m_F=\pm 1$ Zeeman sublevels, denoted as $|g_{\pm1} \rangle$, to the $F'=0, \; m'_F=0$ excited state $|e\rangle$, as indicated in Fig.~\ref{fig:overview}. The $F=1,\; m_F=0$ sublevel of the ground state, is denoted as $|g_0\rangle$. The transition is driven by weak vector vortex probe light (\ref{eq:Efield}) in the presence of a static magnetic field (\ref{eq:Bfield}) with arbitrary inclination, $\polarB$, and azimuth, $\azimuthB$.
 
The Hamiltonian in the Zeeman basis reveals a strong relationship between the
geometry of the applied field and the energy of the system: 
\begin{equation}
\begin{aligned}
\label{eq:hamiltonian-zeeman}
    \hamiltonian[Z]  & = \hbar \Big[ \pm\Omega_\|  \dyad{g_{\pm 1}}  - \e{\mp \azimuthB} \frac{\Omega_\perp}{\sqrt{2}} \dyad{g_{\pm1}}{0} \\
     &  -\frac{\Omega_{\pm}}{2} \dyad{g_{\pm 1}}{e} \Big] + \hc,
\end{aligned}
\end{equation}
where we have assumed resonant optical coupling. 
Here $\Omega_{\pm}=\exp(\mp i \ell \phi) \Omega_R/\sqrt{6}$ denotes the optical coupling, where $\Omega_R$ is the Rabi frequency and we have considered the appropriate Wigner-Eckart coefficients. 
The effect of the magnetic field component along and orthogonal to the optical axis imposes a Zeeman shift on the states $|g_{\pm 1}\rangle$, and mixing of the Zeeman sublevels, characterized by $\Omega_\| = \Omega_L \cos\polarB$ and $\Omega_\perp =  \Omega_L \sin \polarB$, respectively, where $\Omega_L=g_{\rm F} \mu_{\rm B} B_{\rm 0}$ is the Larmor frequency.

The Hamiltonian can then be rewritten in terms of spatially dependent partially dressed states $\ket{\wavF[i]}$, such that:
\begin{equation}
\begin{aligned}
\label{eq:hamiltonian_trans}
\hat{H}_\psi & = \hlf[\hbar] \Big[ \larmor \cos\oam \sin\polarB \dyad{\wavF[1]}{\wavF[2]} \\
& + \larmor  \normFacAngle \dyad{\wavF[2]}{\wavF[c]} 
+ \frac{\rabi}{2\sqrt{3}} \dyad{\wavF[c]}{e}\Big] + \hc
\end{aligned}
\end{equation}
The states $\ket{\wavF[i]}$ and the normalisation factor, $\normFacAngle$, depend on $\phi, \azimuthB$ and $\polarB$. See Supplemental Material at [URL will be inserted by publisher] for the relevant transformation and expressions.
In the spatially dependent basis, the structure in the optical coherence is now explicitly manifest in the magnetic interaction. In the absence of a magnetic field, \(\ket{\wavF[1]}\) does not interact with the optical fields, being equivalent to the unperturbed ground state \(\ket{0}\), but in the presence of a transverse magnetic field, there are certain values of \(\azimuth\) for which the coherence still necessarily vanishes \emph{i.e.}~for \(\azimuth = n \pi/(2\ell) \forall n\in\naturals_0\), creating a magnetically-induced, spatially dependent dark state, where there can be no absorption once the steady state is reached.

\begin{figure}[tb]
  \centering
  \includegraphics[width=.9\linewidth]{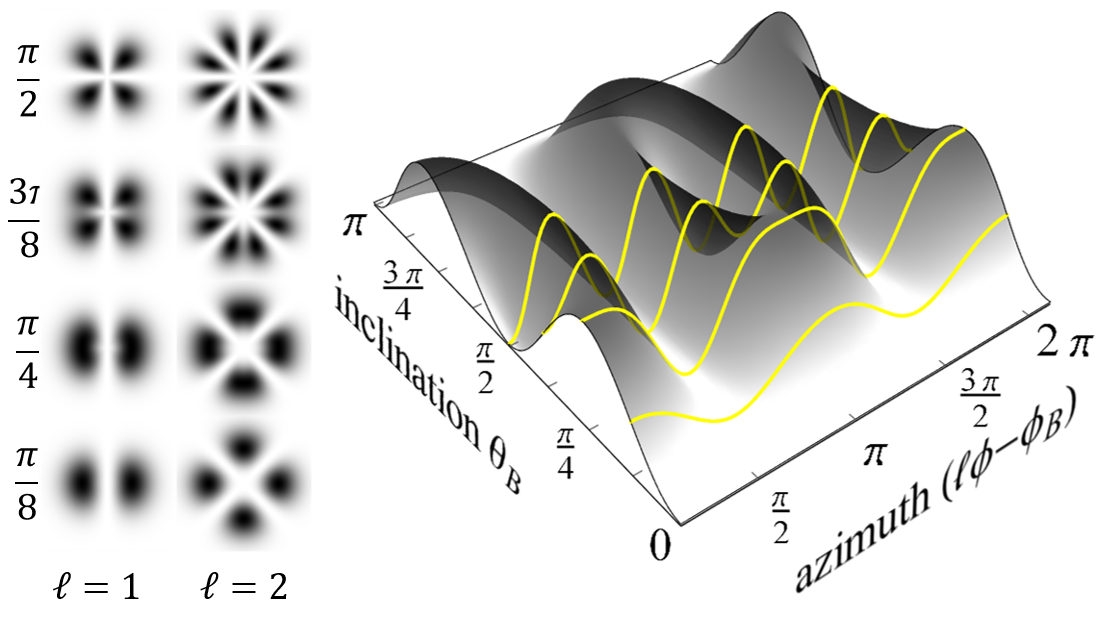}
  \caption{\label{fig:splitting}
   Illustration of the transition probability, $\tp$, based on Fermi’s golden rule. a) Predicted absorption profiles for vector beams with $\ell=1$ and $2$ for the indicated inclination angles. b) $\tp$ as a function of the magnetic field alignment.}
\end{figure}

Using Fermi's golden rule (FGR), and so considering the cumulative probability that a photon will transition between \(\ket{\wavF[1]}\) and \(\ket{\wavF[c]}\), we obtain a concise insight into the analytical form of the interaction:
\begin{equation}
\begin{aligned}
\label{eq:probability}
   \tp \propto\, &  \larmor^4 \rabi^2 \sin^2\polarB \cos^2(\oam - \azimuthB) \\
   &\times  \Big[ \cos^2\polarB + \sin^2\polarB \sin^2(\oam - \azimuthB) \Big].
\end{aligned}
\end{equation}
Rewriting Eq.~(\ref{eq:probability}) as a cosine Fourier series, we can identify the azimuth, $\azimuthB$ and the inclination, $\polarB$, from the phase and magnitude of the Fourier components, as
\begin{align}\label{phiB}\azimuthB & = 2^{-1}\arg(\fourieraz(\tp))_{2\ell},\\ \label{thetaB}
\sin^4\polarB &=\sqrt{8/\pi} \abs{\fourieraz(\tp)}_{4\ell},
\end{align}
forming the basis of what we might call a spatial atomic compass. The transition probability, $\tp$, and selected absorption profiles are shown in Fig.~\ref{fig:splitting}. Rotating $\mathbf{B}$ azimuthally results in a $1/\ell$-fold rotation of the absorption profile, 
whereas its inclination results in a splitting of the absorption pattern. The latter is reminiscent of the splitting of an absorption peak observed in \cite{margalit2013degenerate}.

We will show in the following that the analytic predictions based on FGR agrees qualitatively with our experimental results. It fails, however, to describe some of the subtle atomic response, especially when dealing with $\Bv$ fields that are largely orthogonal to the optical propagation direction, or for higher probe power. A rigorous treatment, based on optical Bloch equations \cite{clark2016sculpting,hamedi2018,sharma2017phase}, results in simulations which are in excellent quantitative agreement with our measurements, however without permitting a simple analytical description. See Supplemental Material at [URL will be inserted by publisher] for an overview.

\begin{figure}[bt]
\centering
\includegraphics[width=\linewidth]{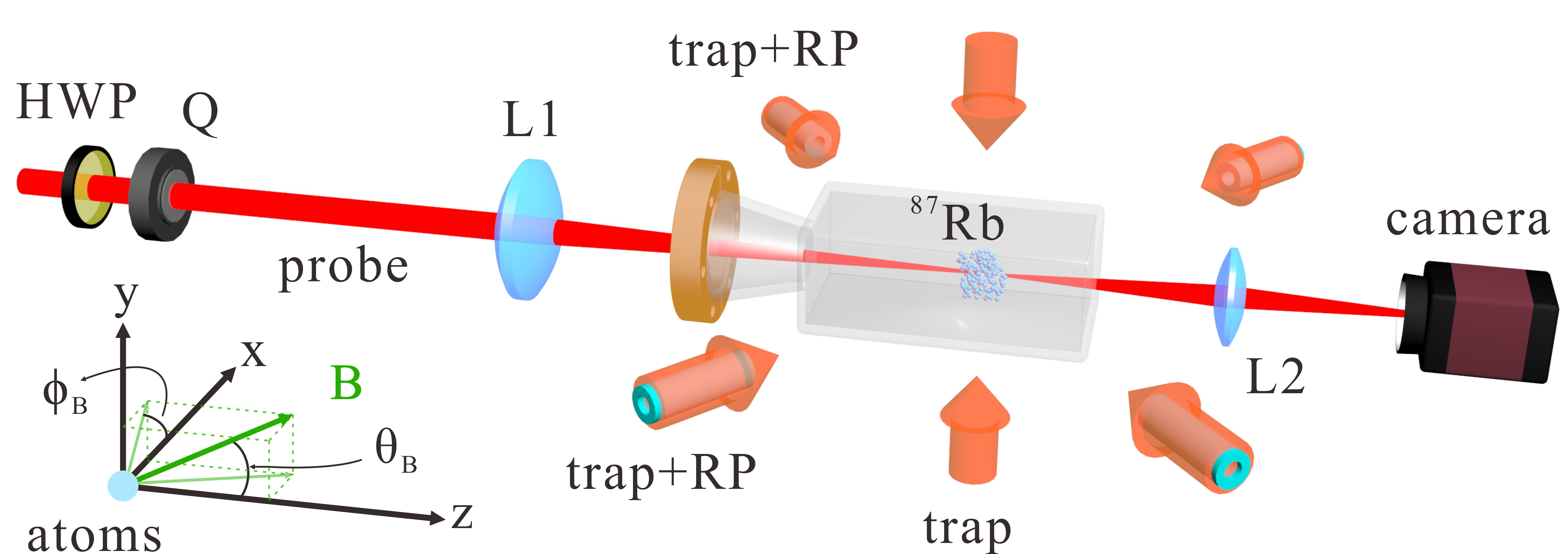}
  \caption{\label{fig:setup}
Schematic diagram of the experimental geometry. The atoms are in the far field of the q-plate (Q), and are imaged to the camera plane. The optical pumping configuration for the SpOT 
is explained in \cite{radwell2013cold}. The bottom left inset shows the defining coordinate system and the alignment of the magnetic field. HWP: half-wave plate, L: lens, RP: repump.}
\end{figure}

\begin{figure*}[tbh]
  \centering
  \includegraphics[width=0.9\linewidth]{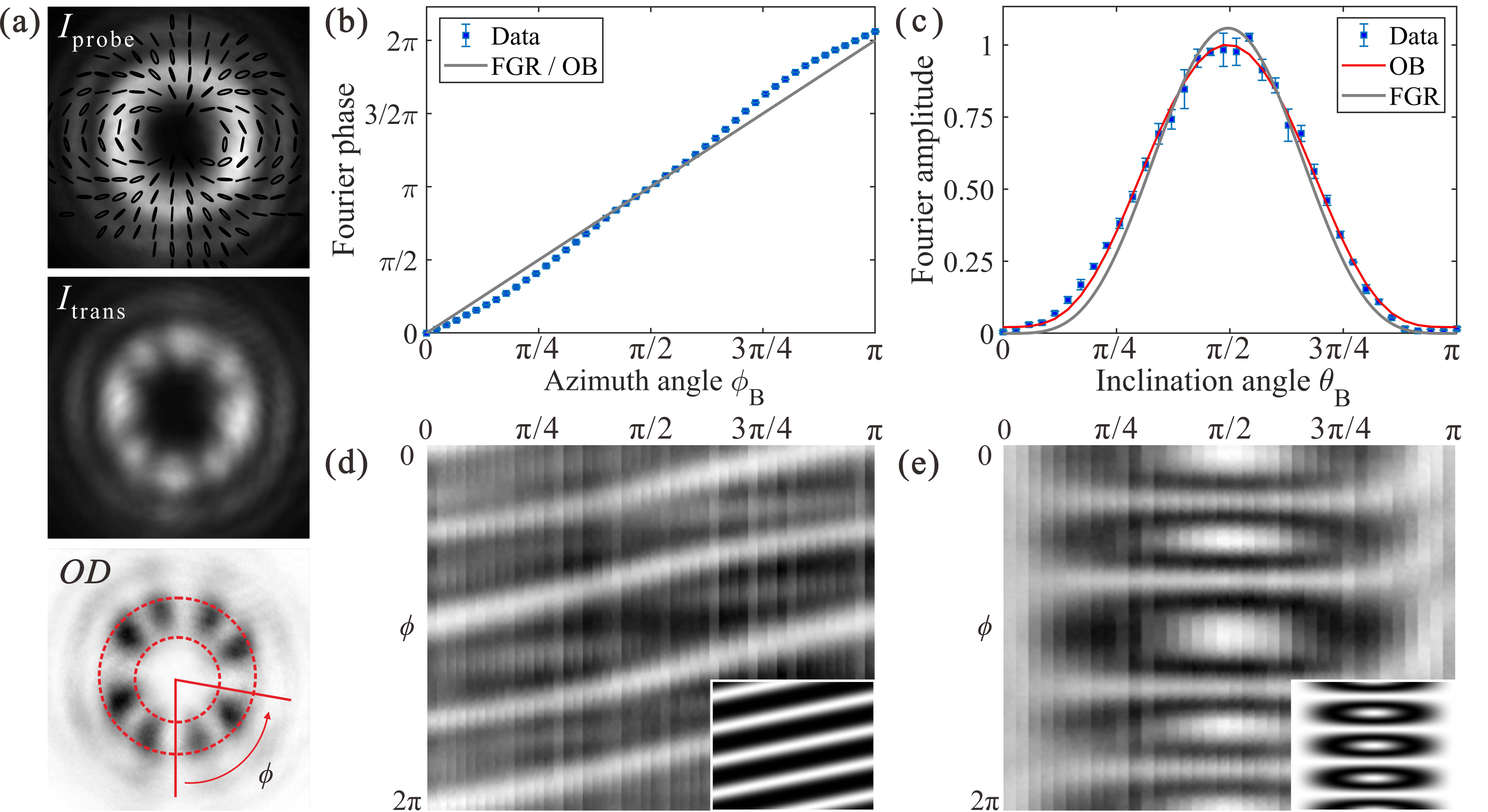}
  \caption{Magnetic field alignment from Fourier analysis of the atomic absorption profiles. (a): example images ($600\times 600\; \mu{\rm m}^2$) of the probe intensity, $I_{\rm probe}$, transmitted light, $I_{\rm trans}$, and resulting absorption profile, ${\rm OD}$, with the analysis region indicated in red. (b)-(c): Dependence of the $2\ell$ and $4\ell$ Fourier components on $\azimuthB$ and $\polarB$ of the $\mathbf{B}$, and comparison with FGR and optical Bloch model (OB). Error bars of the data points (blue) represent the standard deviation of 3 or 5 runs. (d)-(e): corresponding compilations of the unwrapped OD images for steps of 70 and 87 mrad 
  respectively, with the FGR prediction as insets.
  }
    \label{fig:experimental-results}
\end{figure*}

\paragraph{Experimental realization and discussion:}
A cold atomic cloud, optical probe light and a global magnetic field were created and combined in a simple linear arrangement (Fig.~\ref{fig:setup}), setting various values of $\Bv$. For each alignment, the spatially dependent absorption profile, proportional to the optical density
$ {\rm OD} = \ln[(I_{\rm probe} - I_{\rm back})/(I_{\rm trans} - I_{\rm back})]$,
was recorded and the consequent Fourier components extracted,
where $I_{\rm probe}$, $I_{\rm trans}$ and $I_{\rm back}$ represent the intensity of the probe before and after absorption, and the background intensity, respectively, with examples shown in Fig.~\ref{fig:experimental-results}a).
The atomic cloud was formed from $^{87}$Rb atoms collected in a magneto-optical trap (MOT), before transfer to the $F=1$ ground state of a dark spontaneous-force optical trap (SpOT) \cite{radwell2013cold} (Fig.~\ref{fig:overview}). Approximately $5\times 10^{7}$ atoms were evenly distributed over the three Zeeman sub-levels, while maintaining an atomic density of $10^{11}$ cm$^{-3}$ and a temperature of $100\ \mu$K. The trapping, repump and depump beams, as well as the MOT's magnetic quadrupole field, were then switched off and the cloud expanded freely for $3.5$~ms before interaction with the vector vortex probe light.
Such light, locked to the $F=1 \to F'=0$ transition, was generated with a q-plate \cite{marrucci2006optical}, where the measured polarisation and intensity profile is shown in Fig.~\ref{fig:experimental-results}(a) for $\ell=2$. The probe power was varied over a range from $0.03\;\mu$W to $0.5\;\mu$W, but had greatest agreement with Eq. (\ref{eq:probability}) for lower values, corresponding to a perturbative regime. The results presented in Fig.~\ref{fig:experimental-results} were taken with a total beam power of $0.13$ $\mu$W, corresponding to a Rabi frequency of $\Omega_R$=$2\pi\times0.26$ MHz in the region of interest, indicated by the red lines in Fig.~\ref{fig:experimental-results}(a).

Before interaction, a global $\mathbf{B}$ field with a fixed magnitude of $10^{-4}$ T was generated and applied using three orthogonal sets of rectangular coils, varying $\azimuthB$ and $\polarB$ in steps of $80\;$mrad for each run. Following standard practice, the desired external field, $\Bv$, was added to a cancellation field, already applied during the operation of the MOT and SpOT, and opposing any spurious environmental fields at the position of the atomic cloud. 

Qualitatively, the results confirm that the absorption pattern rotates azimuthally with applied $\Bv$, Fig.~\ref{fig:experimental-results}(d), and splits from $2\ell$ to $4\ell$ lobes when its inclination from the optical axis rises from 0 to $\pi/2$, Fig.~\ref{fig:experimental-results}(e). The quantitative comparisons, based on Fourier analysis of the data and models, are presented in Fig.~\ref{fig:experimental-results}(b) and (c). The analytical predictions of Eq.~(\ref{phiB}) and (\ref{thetaB}), shown as grey lines, are largely in agreement with the data. 
In our experiment, we had to balance the low probe intensity required for the weak-coupling limit with the necessity for high contrast absorption images from our detectors. Therefore, although providing a concise insight, the perturbative regime required for equation \eqref{eq:probability} was not fully applicable to our conditions, and we may observe a spatial analogue of intensity broadening. A model based on the full optical Bloch equations, normalised to the data and fitting on the beam intensity, leads to excellent agreement. 

The remaining discrepancies in the data are likely technical in origin. The polarization profile of our probe shows small ($\phi$-dependent) degrees of ellipticity, corresponding to an imbalance between the $\sigma_\pm$ light components. Furthermore, we are using the magnetic field cancellation coils of a standard MOT setup to define our $\mathbf{B}$ alignment, and any incomplete cancellation of environmental fields may result in 
a small tilt from the desired alignment, providing a likely source of systematic experimental uncertainty. Random error however, was reduced to acceptable levels, as indicated by standard deviations, averaging over five and three runs for each $\polarB$ and $\azimuthB$, respectively (Fig.~\ref{fig:experimental-results}(b)-(c)). The corresponding precision of the $\mathbf{B}$ field alignment, after inverting Eqs.~\ref{phiB},\ref{thetaB} was $30\;$mrad for both $\azimuthB$ and $\polarB.$  

\paragraph{Conclusions:}
\label{sec:conclusions}
Throughout, we exposed a spatial relationship between magnetic field
alignment and phase-shaped light on interaction with an atomic cloud. Using
this relationship, we have shown, analytically and experimentally, how an atomic
cloud may be used as a three-dimensional compass, without explicitly invoking
time-dependent effects. The 3D-information is derived from   individual absorption images obtained in single-axis optical probing, where a vector vortex probe beam both generates and measures the atomic polarization. These results hold in the steady-state limit and can be largely independent of applied field strength, offering opportunities for a new branch of magnetic sensing. This parallel geometry is promising for the development of chip-based and miniature sensors.
Although so far we have only considered simple absorption patterns, this spatial mapping to a magnetic field can be quite general: providing an original tool-kit for the spatial manipulation of magnetic dipole and quadrupole moments in atoms. With asymmetric polarization patterns we not only expect greater field information in the absorption patterns, but we would also obtain programmable dispersion relations, where early results suggest practical gains in magneto-optical rotation as well as fundamental insight into spatial analogues of the Kramers-Kronig relations.

\begin{acknowledgments}
The authors would like to thank Péter Domokos for supporting the collaboration and Gergely Szirmai for stimulating discussions regarding Fourier
transforms in unusual bases.
FC and SF-A acknowledge financial support
from the European Training Network ColOpt, funded by the European Union Horizon 2020 program
under the Marie Sklodowska-Curie Action, Grant Agreement No. 721465.
TWC acknowledges support by the National Research, Development and Innovation Office of Hungary (NKFIH) within the Quantum Technology National Excellence
Program (Project No. 2017-1.2.1-NKP-2017-00001). JW acknowledges support by the China Scholarship Council (CSC) (No. 201906280228).
\end{acknowledgments}
\section{References}
\bibliography{main}
\newpage

\onecolumngrid

\section{Supplemental material:}
Here we present some details of the underlying theoretical calculations related to the absorption images obtained via our model based on Fermi's golden rule and optical Bloch equations. This includes the state transformations from atomic to partially dressed states and the subsequent extraction of magnetic alignment.

\section{Fermi's golden rule model (FGR)}
\label{sec:fermiModel}

\subsection{Transformation}
\label{sec:transformation}
In this section we describe the transformation from the Hamiltonian in terms of the atomic states, Eq.(4), to that in terms of partially dressed states, Eq.(5).

When considering a system with a transverse magnetic field,
there are two conventionally used frames of reference, as determined by the orientation of the quantization axis. Either the quantization axis is parallel to the axis of propagation, such as to simplify the decomposition of the light, or it is aligned with the direction of the net magnetic field, as to remove the interaction between the new ground levels. In this work however, we consider a separate approach,
where the excited state interacts with a coherent superposition of spatially
dependent ground states, forming partially dressed states. In this frame, the Hamiltonian is reduced to three
interactions, similarly to the combined field basis. The Zeeman splitting is removed from the ground-state energy levels, and the total interaction can be described as a four-step ladder system.

The transformation states are then given by
\begin{equation}
\begin{aligned}
  \ket{\wavF[1]} \equiv &-\frac{\ii \e{-\ioam} \sin\polarB \sin(\oam - \azimuthB)}{\sqrt{2} \normFacAngle} \ket{g_{+1}} + \frac{\cos\polarB}{\normFacAngle}\ket{0} 
-\frac{\ii \e{+\ioam} \sin{\polarB} \sin(\oam - \azimuthB)}{\sqrt{2} \normFacAngle} \ket{g_{-1}},\\
\ket{\wavF[2]} \equiv &- \frac{\e{-\ioam} \cos\polarB}{\sqrt{2}\normFacAngle} \ket{g_{+1}} + \frac{\ii\sin\polarB \sin(\oam-\azimuthB)}{\normFacAngle} \ket{0}
 - \frac{\e{\ioam}\cos\polarB}{\sqrt{2}\normFacAngle} \ket{g_{-1}}\textrm{ and }\\
\ket{\wavF[c]} \equiv &-\ort[\e{- \ioam}] \ket{g_{+1}} + \ort[\e{\ioam}] \ket{g_{-1}},
%
\end{aligned}
\end{equation}
where \(\normFacAngle = \sqrt{\cos^2\polarB + \sin^2\polarB\sin^2(\oam -
  \polarB)}\) and the excited state, $\ket{e}$, is unchanged. We also note that the transformation, \(\unitary =
\{\ket{\wavF[1]},\ket{\wavF[2]},\ket{\wavF[c]},\ket{\wavF[e]}\}\), was applied to the Zeeman Hamiltonian before the rotating wave approximation, under conventional constraints.

\subsection{Spatial splitting}
\label{sec:Splitting}
In 2013, Margalit, Rosenbluh and Wilson-Gordon, \cite{margalit2013degenerate} in the context of frequencies, showed that it was
possible to split an absorption peak in an \(F_{g}= 1 → F_{e }= 0\) transition.
This was something of a surprise, as such frequency splittings had been attributed to the creation of high-order
ground states and thus were not deemed possible for excitations
between transitions with lower total angular momentum. Margalit \emph{et al.}
however, showed that for their (and this) system, it was now possible to independently measure both the value of \(B_{x}\) and the sign and value
of \(B_{z}\). Importantly however, the rending of their absorption peak
was a function of magnetic field strength, not space.

Spatial splitting also reveals magnetic information. Using our partially dressed states (outlined above), we can find the transition between states $\ket{\wavF[1]}$ and  $\ket{e}$ from Fermi's golden rule. The dependence on magnetic alignment can be seen from the transmission probability, and immediately extracted from the associated Fourier series:

\begin{equation}
\begin{aligned}
    \tp \propto&   \frac{2\pi}{\hbar} |\langle \wavF[1]|\hat{H}_\psi|e\rangle|^2\\
    =& \frac{2\pi}{\hbar} \left (\frac{\hbar}{2}\right)^6 |\larmor \cos\oam \sin\polarB|^2 |\larmor  \normFacAngle|^2 \left|\frac{\rabi}{2\sqrt{3}}\right|^2\\
  = \, &\larmor^4 \rabi^2 \Big[ \frac{1}{2}\sin^2\polarB - \frac{3}{8} \sin^4\polarB    + \frac{1}{2}\Big( \sin^2\polarB-\sin^4 \polarB\Big)\cdot\cos(2\oam -2\azimuthB)    +\Big. \frac{1}{8} \sin^4\polarB \cdot\cos(4 \oam - 4 \azimuthB)\Big],
\end{aligned}
\end{equation}
where the last expression is simply a matter of trigonometric identities.

\section{Optical Bloch equation model (OB)}
The transmission probability (above) concisely captures the main physics and provides an intuitive insight into the interaction. It is limited in its applicability however, as Fermi's golden rule is a perturbative approximation not suitable to strong interaction or appreciable dissipation. Here we consider the appropriate optical (Maxwell-)Bloch equations for our experimental system.

Our Bloch equations were constructed from the Lindblad master equation \cite{clark2016sculpting,hamedi2018,sharma2017phase},
\[
  \dv{t}\adm = -\frac{\ii}{\hbar}[\hamiltonian, \adm] - \hlf \left(\decayop\adm + \adm\decayop\right) + \repopop,
\]
where we have separated the Liouville operator in terms of the relaxation, $\decayop$, and repopulation, $\repopop$.  These were defined for our specific states:
\begin{align}
  \label{eq:1}
  \decayop &= \sum_{i}\gamma\dyad{g_{i}} + (\gamma + \decay)\dyad{e} \text{ and  }\\
  \repopop &= \sum_{i}\frac{1}{3}(\gamma + \decay \adm[e,e]).
\end{align}

The optical Bloch equations were then defined using the Zeeman-basis Hamiltonian, under the rotating wave approximation, outlined in the original letter ($\hamiltonian[Z]$).

Under the FGR model, absorption was qualitatively associated with the transition probability through the ladder system. For the optical Bloch model however, absorption was considered more precisely: as the relative change in electric field following propagation through the atomic cloud.

Although the form for unstructured light is well known, the presence of phase-structured light complicates the electric-field propagation, such that both real and imaginary components of the density operator contribute. The resulting relationship can be expressed by
\begin{equation}
  \label{eq:2}
  \frac{1}{\Eamp_{\pm }k}\pdv{z}\Eamp_{\pm }=2\sqrt{3}\pi\numberDensity z \frac{\decay}{\Omega_{R\pm}\energy_{e}^{2}}\left[ \cos(\varphi_{\pm})\textrm{Im}(\hat{\rho}_{\mp,e}) + \sin(\varphi_{\pm})\textrm{Re}(\hat{\rho}_{\mp,e}) \right],
\end{equation}
where $\pm$ labels the transitions between the $g_{\pm 1}$ ground states and the excited state, e; $\Eamp$ is the field amplitude; $\varphi$ is the associated phase; $\rabi$ is the Rabi frequency; $k$ is the wavenumber; $\numberDensity$ is the number of atoms per $cm^{3}$; $\energy_{e}$ is the excited state energy and $z$ is the propagation distance through the cloud.
\end{document}